\documentclass[aps,prl,twocolumn,superscriptaddress]{revtex4-1}
\usepackage{graphicx}
\usepackage{amsmath}
\usepackage{nccmath}
\usepackage{amsfonts} 
\usepackage{pifont}
\usepackage{color}
\newcommand{\cmark}{\ding{51}}%
\newcommand{\xmark}{\ding{55}}%

\newcommand*{\citen}[1]{%
  \begingroup
    \romannumeral-`\x 
    \setcitestyle{numbers}%
    \cite{#1}%
  \endgroup   
}

\begin{document}
\title
{Broadband mixing of ${\cal PT}$-symmetric and ${\cal PT}$-broken phases in photonic heterostructures with 
a one-dimensional loss/gain bilayer} 
\author{Ege \"Ozg\"un} 
\email{ozgune@bilkent.edu.tr}
\affiliation{NANOTAM-Nanotechnology Research Center, Bilkent University, 06800 Ankara, Turkey}
\author{Andriy E. Serebryannikov} 
\email{aserebry@esat.kuleuven.be}
\affiliation{Faculty of Physics, Adam Mickiewicz University, 61-614 Pozna\'{n}, Poland} 
\author{Ekmel Ozbay} 
\affiliation{NANOTAM-Nanotechnology Research Center, Bilkent University, 06800 Ankara, Turkey} 
\affiliation{Department of Physics, Department of Electrical and Electronics Engineering and UNAM-Institute of Materials Science and Nanotechnology, Bilkent University, 06800 Ankara, Turkey}
\author{Costas M. Soukoulis} 
\affiliation{Institute of Electronic Structure and Laser, FORTH, 71110 Heraklion, Crete, Greece} 
\affiliation{Ames Laboratory and Department of Physics and Astronomy, Iowa State University, Ames, Iowa 50011, USA}

\begin{abstract}
Combining loss and gain components in one photonic heterostructure opens a new route to efficient manipulation by radiation, transmission, absorption, and scattering of electromagnetic waves. Therefore, loss/gain structures enabling ${\cal PT}$-symmetric and ${\cal PT}$-broken phases for eigenvalues have extensively been studied in the last decade. In particular, translation from one phase to another, which occurs at the critical point in the two-channel structures with one-dimensional loss/gain components, is often associated with one-way transmission. In this report, broadband mixing of the ${\cal PT}$-symmetric and ${\cal PT}$-broken phases for eigenvalues is theoretically demonstrated in heterostructures with four channels obtained by combining a one-dimensional loss/gain bilayer and one or two thin polarization-converting components (PCCs). The broadband phase mixing in the four-channel case is expected to yield advanced transmission and absorption regimes. Various configurations are analyzed, which are distinguished in symmetry properties and polarization conversion regime of PCCs. The conditions necessary for phase mixing are discussed. The simplest two-component configurations with broadband mixing are found, as well as the more complex three-component configurations wherein symmetric and broken sets are not yet mixed and appear in the neighbouring frequency ranges. Peculiarities of eigenvalue behaviour are considered for different permittivity ranges of loss/gain medium, i.e., from epsilon-near-zero to high-epsilon regime.       
\end{abstract}
       
\maketitle
Being a weaker condition compared to hermiticity, ${\cal PT}$-symmetry can still yield real positive eigenvalues for Schr\"odinger's equation \cite{qm1,qm2,qm3}. The investigations of ${\cal PT}$-symmetry 
have not been restricted to the area of quantum mechanics. 
The realization of ${\cal PT}$-symmetric structures in optics has been suggested in the 2000s. In Ref. \citen{wg5}, a parallel-plate waveguide with 
${\cal PT}$-symmetric interior has been proposed. A large portion of the optics relevant theoretical studies has been focused on one-dimensional lattices with gradual variation of the refractive index \cite{wg1a,wg1b,wg2a}. 
Moreover, ${\cal PT}$-symmetry has been experimentally observed in optical lattices \cite{wg3,wg4}.  
Later,   
conservation relations and ${\cal PT}$-transition properties of one-dimensional photonic heterostructures \cite{GCS_PRL,GCS} have been studied. 
In this case, the eigenvalues of the $\mathbb{S}$-matrix are unimodular and flux conserving in ${\cal PT}$-symmetric phase. In ${\cal PT}$-broken phase, they have different magnitudes, one of which is larger and the remaining one is smaller than unity \cite{GCS}. They are known to correspond to amplification and attenuation, respectively, 
and satisfy the generalized conservation condition \cite{GCS_PRL,GCS}. 
The two above-mentioned phases are separated by the spontaneous ${\cal PT}$-symmetry breaking point (called critical, or exceptional, or phase 
transition point), at which $\mathbb{S}$-matrix has only one eigenvalue instead of the two ones. In this regime, the system can be fully transparent for the light incident from one side of the structure while reflectivity is enhanced for the opposite-side incidence \cite{UD-mix}. 
The role of the critical point for achieving unidirectional transmission has also been highlighted in many other theoretical and experimental studies, 
e.g., see Refs. \citen{asym-PLA,IEEE-08,except,Scherer}. Unidirectional invisibility and cloaking in the structures having ${\cal PT}$-symmetric components have been studied in detail \cite{UD-mix,cloak,OL-cloak}. 
Moreover, lenses and cavities with ${\cal PT}$-symmetry should be mentioned \cite{AA-JOPT,PTWGM,PTCav}. 
While bilayers, multilayers, and waveguides with a finite-extent cross section were commonly considered in the studies of 
optical structures with ${\cal PT}$-symmetry, the attention has also been attracted by two-dimensional and quasiplanar one-dimentional arrays \cite{PRA-EO,oneway,Feng,exp-NatCom}. 
Specifics of the epsilon-near-zero (ENZ) range of permittivity has been intestigated for the bilayer structures \cite{Alu,ENZ2015}. In particular, 
conditions of tunneling through ${\cal PT}$-symmetric ENZ bilayers have been considered \cite{Alu}. 
Some exotic regimes like loss-induced superscattering and gain-induced absorption \cite{Feng} are also worth mentioning. 
At the same time, new opportunities are promised by recent advances in full control of electromagnetic waves with the aid of metasurfaces \cite{msf01,msf02,msf04,msf05,msf07}. 
In particular, some extreme regimes of polarization conversion have recently been demonstrated \cite{msf01,polconv02,polconv04a,polconv04b}, whereas some others are expected to be achieved soon. This opens a new route 
to multichannel structures with ${\cal PT}$-related properties, which do not need two-dimensional loss/gain components. 
Indeed, although the principal possibility of co-existence of  ${\cal PT}$-symmetric and ${\cal PT}$-broken phases is a known feature for the case of physical dimensions higher than one (and, hence, more than two input and output channels may exist)\cite{GCS}, there is high demand in simpler structures  
that would enable easily realizable scenarios.  

In this report, we theoretically demonstrate the co-existence of ${\cal PT}$-symmetric and ${\cal PT}$-broken sets of eigenvalues of the $\mathbb{S}$-matrix in a four-channel photonic heterostructure with a one-dimensional bilayer, 
while additional two channels required for the phase mixing are created due to thin polarization converting componenets (PCCs) that are placed at the end-faces. 
Our main goal is to show the principal possibility and find the basic features of this regime.  
Therefore, in the contrast with a two-dimensional case of loss/gain media, the function of the creation of additional transmission channels and that of combining gain and loss are $separated$ in space, so far as they are performed by the different components of the entire structure. 
There are two input and two output channels (ports) in the studied heterostructures, whereas each of the PCCs is assumed to be passive, lossless, and quasiplanar. 
To the best of our knowledge, the possibility of the simultaneous existence of ${\cal PT}$-symmetric and ${\cal PT}$-broken phases in a photonic system with one-dimensional loss/gain component has not yet been discussed in the literature. 
We refer to the co-existence of ${\cal PT}$-symmetric and ${\cal PT}$-broken phases in the same frequency range as the phase mixing. The goals of this study include the validation of the suggested approach, finding and comparison of different configurations enabling the mixing of symmetric and symmetry-broken phases, and the formulation of the conditions providing the mixing at the minimal number of the structural components. These goals can be achieved by using 
the standard $\mathbb{S}$-matrix formalism, while advanced techniques are not required.  
We investigate a wide range of variation of permittivity of the loss and gain components. Among others, it includes the ENZ region, which is known for such exotic properties as squeezing, supercoupling,  
stretching of wavelength, enhancing nonreciprocity, and time-reversal symmetry breaking \cite{ENZ-sci}.
Here, the study of the ENZ range and consideration of PCCs are restricted according to the goals of this paper.  
Design of PCCs is beyond its scope.  

\section*{Results}
\textbf{General Model.}
\begin{figure} [t]
\centerline{\includegraphics[scale=0.5]{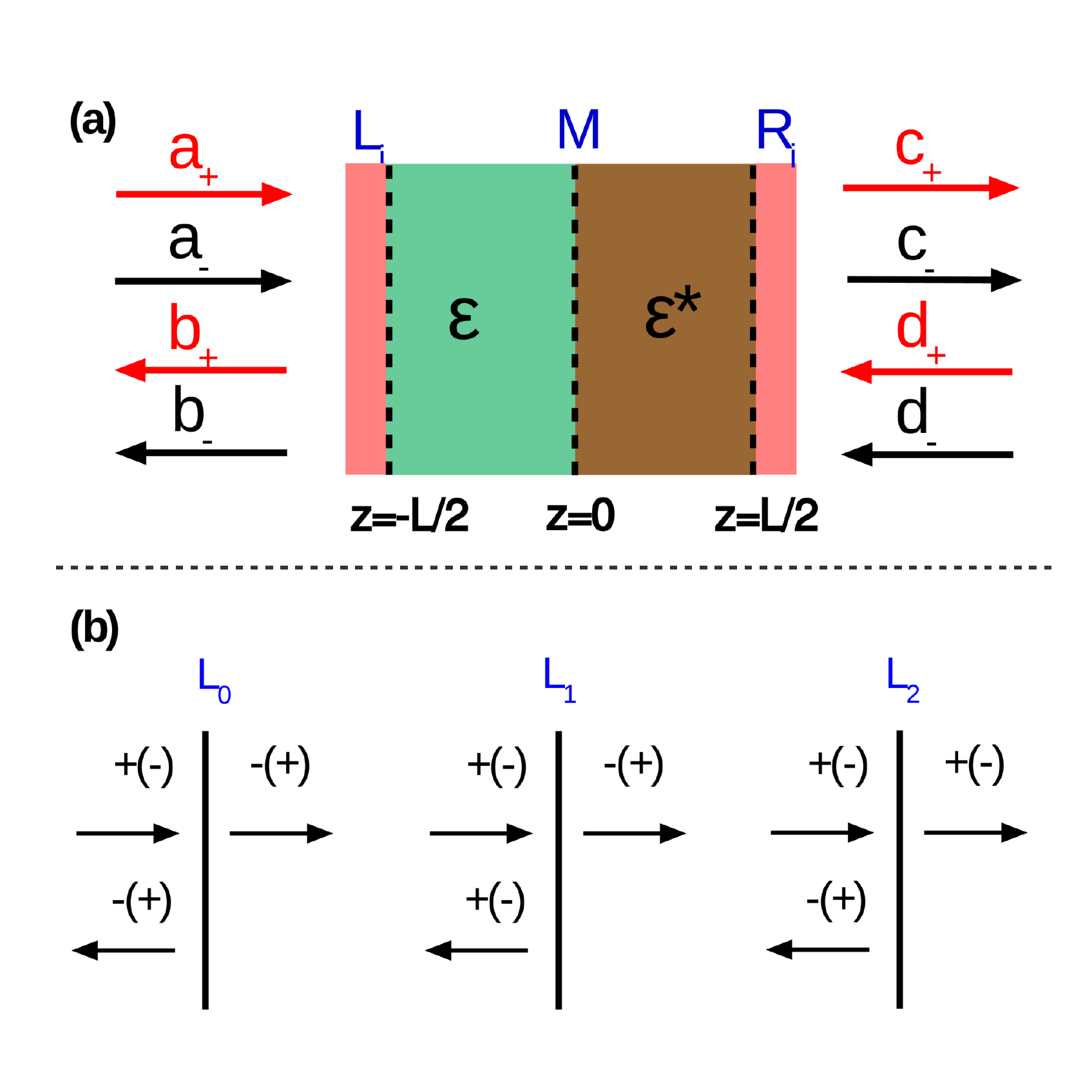}}
\caption{(Color online) (a) Schematic view of the studied heterostructure. There are four channels in total, which contain two input and two output channels corresponding to the right and left circular polarizations (CPs), which are described by eight coefficients for the scattering amplitudes, $a_{\pm}$, $b_{\pm}$, $c_{\pm}$, $d_{\pm}$. The middle region labeled by $M$ consists of gain and loss components, 
satisfying $\varepsilon(z)=\varepsilon^{\ast}(-z)$. Different types of PCCs (labeled as $L_i$ and $R_i$) can be added at the end-faces of region $M$ to extend a variety of achievable scenarios. (b) Three types of PCCs are considered: $L_0$ component changes the polarization of both the transmitted and reflected waves (perfect polarization conversion from right/left CP to left/right CP is assumed), whereas $L_1$ and $L_2$ change only polarization of the transmitted and only polarization of the reflected waves, respectively. For right-side incidence, the components $R_0$, $R_1$ and $R_2$ show the same properties as their left-side counterparts, i.e., $L_0$, $L_1$, and $L_2$. }
\label{htr}
\end{figure}
The studied photonic heterostructures are assumed to have in the general case the following three components: The loss/gain bilayer and two PCCs, one at each of its end-faces. Capability in polarization conversion is directly connected with the possibility of the creation of additional channels required for obtaining a four-channel configuration. 
Different configurations can be accessed through adding or removing 
one or two PCCs.
The general schematic of the structure is shown in Fig. \ref{htr}(a). The region $M$ is the loss/gain medium that 
satisfies the symmetry condition $\varepsilon(z)=\varepsilon^{\ast}(-z)$.
It is noteworthy that this condition is necessary but not sufficient for ${\cal PT}$-symmetry. 
For the purposes of our study, we assume that PCCs are infinitesimally thin and 
capable of converting right/left circular polarization (CP) to left/right CP perfectly. 
Here, we consider three different types of PCCs. The first-type PCCs denoted by $L_0$ and $R_0$ change the polarization state for both reflected and transmitted waves. 
The second-type PCCs (denoted by $L_1$ and $R_1$) only change the transmitted waves' polarization, whereas the third-type PCCs (denoted $L_2$ and $R_2$) only change the reflected waves' polarization. The properties of PCCs 
of the three types are illustrated in Fig. \ref{htr}(b).
Polarization state cannot be changed by using only bilayer. All of the studied structures are free-standing structures adjusted with the vacuum half-spaces.  

\par 
We build our formalism for a general 1D photonic heterostructure that has four channels, two input and two output, one allowing right circularly polarized light and the other allowing left 
circularly polarized light to pass through. We start from casting the most general expression for the electric field along the $z$-direction that consists of right $(+)$ and left $(-)$ polarized waves. For the left-side half-space, $z<-L/2$, and normal incidence, it is given by  

\begin{ceqn}
\begin{align}
{\bf E(z)} = {\bf E}_+(a_+e^{ikn_0z}+b_+e^{-ikn_0z})+{\bf E}_-(a_-e^{ikn_0z}+b_-e^{-ikn_0z}),
\label{eq:Eq1}
\end{align}
\end{ceqn}
where $a_{\pm}$, $b_{\pm}$ are scattering amplitudes, ${\bf E}_+=E_0{\bf \hat{e}_+}$ and ${\bf E}_-=E_0{\bf \hat{e}_-}$, with ${\bf \hat{e}_+} = 1/\sqrt{2} (1,i)$ and ${\bf \hat{e}_-} = 1/\sqrt{2} (1, -i)$, $k=\omega/c$, $\omega$ is angular frequency, $c$ is velocity of electromagnetic wave, $n_0$ is index of refraction.  
To obtain ${\bf E}$ for the right-side half-space, i.e., at $z>L/2$, we replace $a \rightarrow c$ and $b \rightarrow d$ in Eq. (\ref{eq:Eq1}). 
Similar expressions can be used for the bilayer region. 

By exploiting the symmetry properties of the structure, we can cast the $generalized$ $conservation$ $relation$ for the four-channel case. It is expected to be 
and really identical to such a well-known relation in the two-channel case, which is valid everywhere, except for coherent-perfect-absorption laser points, 
see Ref. \citen{GCS}. It is connected with another key relation known as the generalized unitarity relation.   
Starting from the condition $\varepsilon(z)=\varepsilon^{\ast}(-z)$, which also implies ${\bf E}(z)={\bf E}^{\ast}(-z)$, and using the standard  $\mathbb{S}$-matrix formalism, which is commonly used in quantum mechanics and electromagnetic theory, we obtain the conservation relation in the studied four-channel case as follows:

\begin{ceqn}
\begin{align}
\vert T - 1 \vert = \sqrt{R_L R_R},
\end{align}
\end{ceqn}
where $T \equiv \vert t \vert^2$ is the transmittance, $R_L \equiv \vert r_L \vert^2$ and $R_R \equiv \vert r_R \vert^2$ are the reflectances 
at left-side and right-side illumination, with no restrictions imposed on them; $t$ is transmission coefficient, $r_L$ and $r_R$ are reflection coefficients at left-side and right-side illumination. In line with the $\mathbb{S}$-matrix formalism, these coefficients are fully determined by the scattering amplitudes.   
To achieve the purposes of this study, the conventional $\mathbb{S}$-matrix approach can be used for the arbitrary values 
of $t$, $r_L$, and $r_R$. 
\\
\textbf{Various Configurations.}
There are two general cases that are distinguished in terms of the physics they offer. The first one is the case, in which ${\cal PT}$-symmetric and ${\cal PT}$-broken  eigenvalues may not exist simultaneously, and one phase is translated into another at the critical frequency (Case 1). In the second case, the 
co-existence of the symmetric and symmetry-broken eigenvalues is possible, i.e., a $mixed$ $phase$ may occur 
(Case 2). This case is the focus of our study. It will be shown that there are various configurations 
that differ in the end-faces of the region $M$, which can be utilized for accessing Case 1 and Case 2. 
We start from the simplest configuration, and then consider 
more complex configurations by placing one or both of the components $L_i$ and $R_i$ at the end-faces.     
\par
{\bf Case 1:} The simplest configuration to access this case 
is a bilayer enabling ${\cal PT}$-symmetry (like the region $M$), which has been studied in detail in Refs. \citen{GCS_PRL,GCS}. 
For the configuration $M$, $\mathbb{S}$-matrix has two eigenvalues. Let us generalize it by formally adding the polarization related degree of freedom to the system to obtain the $4X4$ $\mathbb{S}$-matrix, i.e., present it in the same form as used throughout the paper for more complex configurations. Since there are no PCCs in this case,
\begin{ceqn}
\begin{align}
\mathbb{S}(\omega) = \begin{pmatrix} r_R & t & 0 & 0 \cr t & r_L & 0 & 0 \cr 0 & 0 & r_R & t \cr 0 & 0 & t & r_L \end{pmatrix}
\label{eq:Eq3}
\end{align}
\end{ceqn}
yields the same set of eigenvalues twice, with the two eigenvalues in a set, which are given by
\begin{ceqn}
\begin{align}
\lambda_{1,2} = \frac{1}{2} \Big \{ (r_R+r_L) \pm \sqrt{(r_R-r_L)^2 +4t^2} \Big \}.
\end{align}
\end{ceqn}
\par
Next, we add the PCCs to the left and right end-faces, which are assumed to be capable of changing the polarization for both reflected and transmitted waves ($L_0$ and $R_0$). The configuration that we have now is $L_0MR_0$, meaning that if the wave is transmitted it retains the initial polarization because of passing through two PCCs. In turn, if it is reflected, 
right/left CP is changed to left/right CP. Now, the two diagonal blocks of the $\mathbb{S}$-matrix in Eq. (\ref{eq:Eq3}) are coupled due to the added PCCs. After some algebra, we obtain      
\begin{ceqn}
\begin{align}
\mathbb{S}(\omega) = \begin{pmatrix} 0 & t & r_R & 0 \cr t & 0 & 0 & r_L \cr r_R & 0 & 0 & t \cr 0 & r_L & t & 0 \end{pmatrix}.
\end{align}
\end{ceqn}
The four eigenvalues for this configuration are given as
\begin{ceqn}
\begin{align}
\lambda_{1-4} = \frac{1}{2} \Big \{ \pm (r_R+r_L) \pm \sqrt{(r_R-r_L)^2 +4t^2} \Big \}.  
\label{eq:Eq6}
\end{align}
\end{ceqn}
\begin{figure} [t]
\centerline{\includegraphics[scale=0.38]{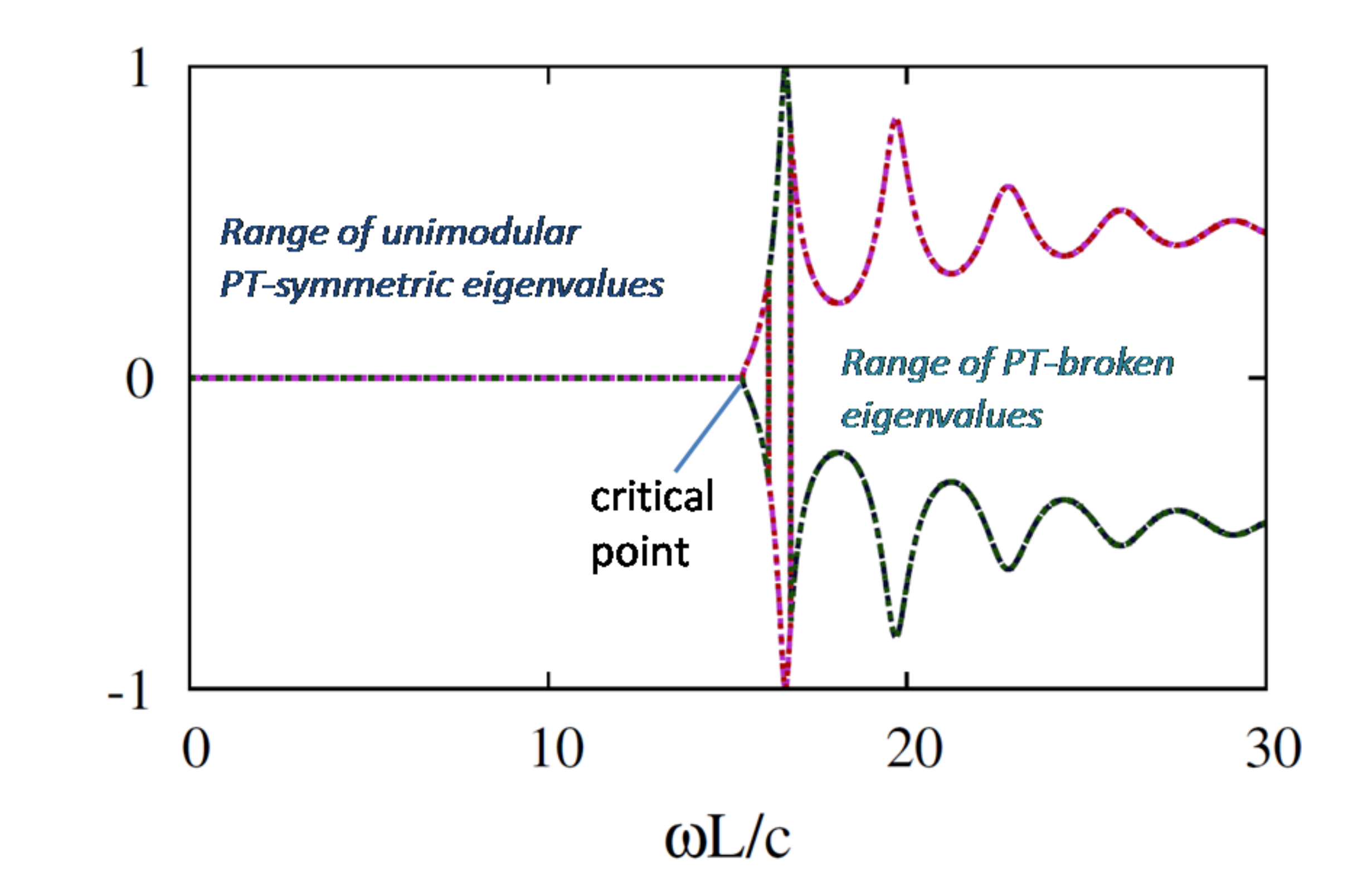}}
\caption{(Color online) Eigenvalues (moduli) in log10 scale vs $\omega{L/c}$ at $\varepsilon = 2 + 0.2i$ in Case 1. For all eigenvalues, the symmetry is spontaneously broken 
at the critical frequency, $\omega_c{L/c}\approx{15}$, and unimodularity is not preserved anymore.} 
\label{2}
\end{figure}
Thus, configurations $M$ and $L_0MR_0$ 
show the same moduli for the eigenvalues and the same basic physics. It is well-known that a $unitary$ $transformation$ preserves the eigenvalues. Hence, all other configurations, for which $\mathbb{S}$-matrices are connected by a unitary transformation with $\mathbb{S}$-matrix of one of two 
above discussed configurations, will have the same eigenvalues. Therefore, these configurations also belong to Case 1, in which the mixed phase for the eigenvalues is not possible. Indeed, they are either ${\cal PT}$-symmetric or ${\cal PT}$-broken at any fixed frequency, except at the critical point.
We can gain access to such configurations by including different combinations of PCCs at one or both end-faces, so 
${\cal PT}$-symmetry is spontaneously broken at the critical point.  
We showed in our study that they include the configurations $L_1MR_1$, $L_2MR_2$, $L_0MR_2$, $L_2MR_0$, $L_1M$, and $MR_1$. 
Here, we do not discuss each of these cases separately, because eigenvalues in all the cases are the same as for 
the configurations $M$ and $L_{0}MR_{0}$, i.e., they are given by Eq. (\ref{eq:Eq6}).
It is interesting that the configurations $L_1M$ and $MR_1$ do not yield the mixed phase despite the fact that polarization conversion is possible at one of the end-faces. 
Since the components $L_1$ and $R_1$ 
only change the polarization of transmitted waves, the      
ability of polarization conversion of the reflected wave is expected to be the $necessary$ (but not sufficient) condition of existence of the mixed phase. 
It is noticeable that the same $\mathbb{S}$-matrices and the same eigenvalues can be obtained 
in Case 1 for the structures with PCCs at two end-faced, with a PCC at one end-face, and without PCCs. 
\begin{figure} [t]
\centerline{\includegraphics[scale=0.85]{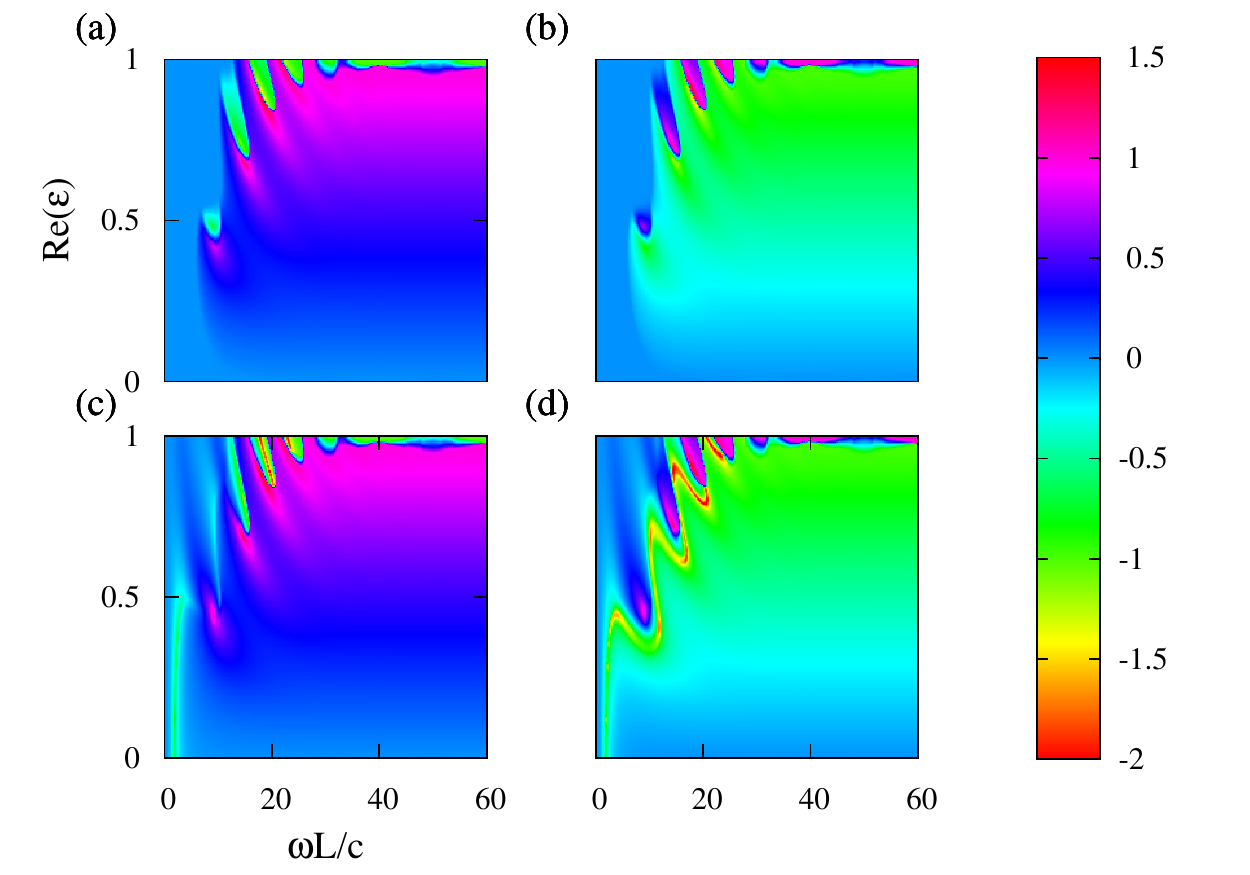}}
\caption{(Color online) The map displaying the magnitudes of the four eigenvalues in log10 scale 
at $Re(\varepsilon)=[0,1]$ and $\omega{L}/c=[0,60]$; $Im(\varepsilon)=0.2$. The upper panels display 
$\lambda_1$ (a) and $\lambda_2$ (b) that are symmetric (unimodular) at $\omega<\omega_c$; 
the lower ones 
display eigenvalues $\lambda_3$ (c) and $\lambda_4$ (d), which are symmetry-broken at any $\omega$. Deviation from unimodular case 
(here - blue color corresponding to 0 at the scale bar) indicates the extent to which the symmetry is broken. Until the value of $\omega_c$ is reached, there is a wide range of $\omega$, where symmetric (unimodular) and symmetry-broken sets of eigenvalues co-exist. After hitting $\omega_c$, the symmetric set also experiences a spontaneous symmetry breaking, and the two eigenvalue sets with broken symmetry start to overlap. \\
}
\label{0-1}
\end{figure}
An example 
is presented in Fig. \ref{2}. 
At $\omega<\omega_c$, all four eigenvalues are $unimodular$ (log10$\lambda_m=0$, $m$=1,2,3,4), being in ${\cal PT}$-symmetric phase and, hence, are flux conserving. At $\omega>\omega_c$, the eigenvalues become reciprocal in two sets, which are in the symmetry-broken phase and, hence, satisfy the generalized conservation relation. These properties are identical to those well-known for one-dimensional structures with two channels. One can see that it is not possible to simultaneously obtain the symmetric and symmetry-broken phases at any fixed value of $\omega{L}/c$. 

{\bf Case 2:} The simplest intuition for accessing this case says that a PCC should be added only to one of the end-faces of the component $M$. 
However, it cannot ensure the phase mixing, as follows from the study of Case 1. 
Thus, let us first add a PCC, specifically $L_0$ to the left end-face, i.e., we now have configuration $L_{0}M$. For the light incident from the left side, polarization of both the reflected and the transmitted wave is assumed to be changeable due to this PCC. If it is incident from the right side, the transmitted wave's 
polarization is still changed, while the reflected wave retains its polarization. These properties are described by the following $\mathbb{S}$-matrix:  
\begin{ceqn}
\begin{align}
\mathbb{S}(\omega) = \begin{pmatrix} r_R & 0 & 0 & t \cr 0 & 0 & t & r_L \cr 0 & t & r_R & 0 \cr t & r_L & 0 & 0 \end{pmatrix}.
\label{eq:Eq7}
\end{align}
\end{ceqn}
This configuration yields the mixed phase for the eigenvalues of the $\mathbb{S}$-matrix, 
so we can obtain symmetric and symmetry-broken sets of eigenvalues at fixed $\omega$. 
The four eigenvalues corresponding to 
Eq. (\ref{eq:Eq7}) are given by  
\begin{subequations}
\begin{ceqn}
\begin{align}
\lambda_{1,2} = \frac{1}{2} \Big \{  (r_R+r_L) \pm \sqrt{(r_R-r_L)^2 +4t^2} \Big \}
\label{eq:Eq8a}, \\
\lambda_{3,4} = \frac{1}{2} \Big \{  (r_R-r_L) \pm \sqrt{(r_R+r_L)^2 +4t^2} \Big \}.
\label{eq:Eq8b}
\end{align}
\end{ceqn}
\end{subequations}
As one set of eigenvalues ($\lambda_{1,2}$) preserves the symmetry and unimodularity 
at $\omega<\omega_c$,
the other set of eigenvalues ($\lambda_{3,4}$) is translated into a symmetry-broken phase in a wide $\omega$-range, even near $\omega{L}/c=0$. 
Once $\omega=\omega_c$ is reached, this symmetry-broken set of the eigenvalues starts to overlap with the 
first set, whose eigenvalues are also in the symmetry-broken phase at that time. Thus, all of the eigenvalues are in the broken phase at $\omega>\omega_c$. 
So, $\omega=\omega_c$ is the boundary between the mixing phase and the all-broken phase cases.
\begin{figure} [t!]
\centerline{\includegraphics[scale=0.85]{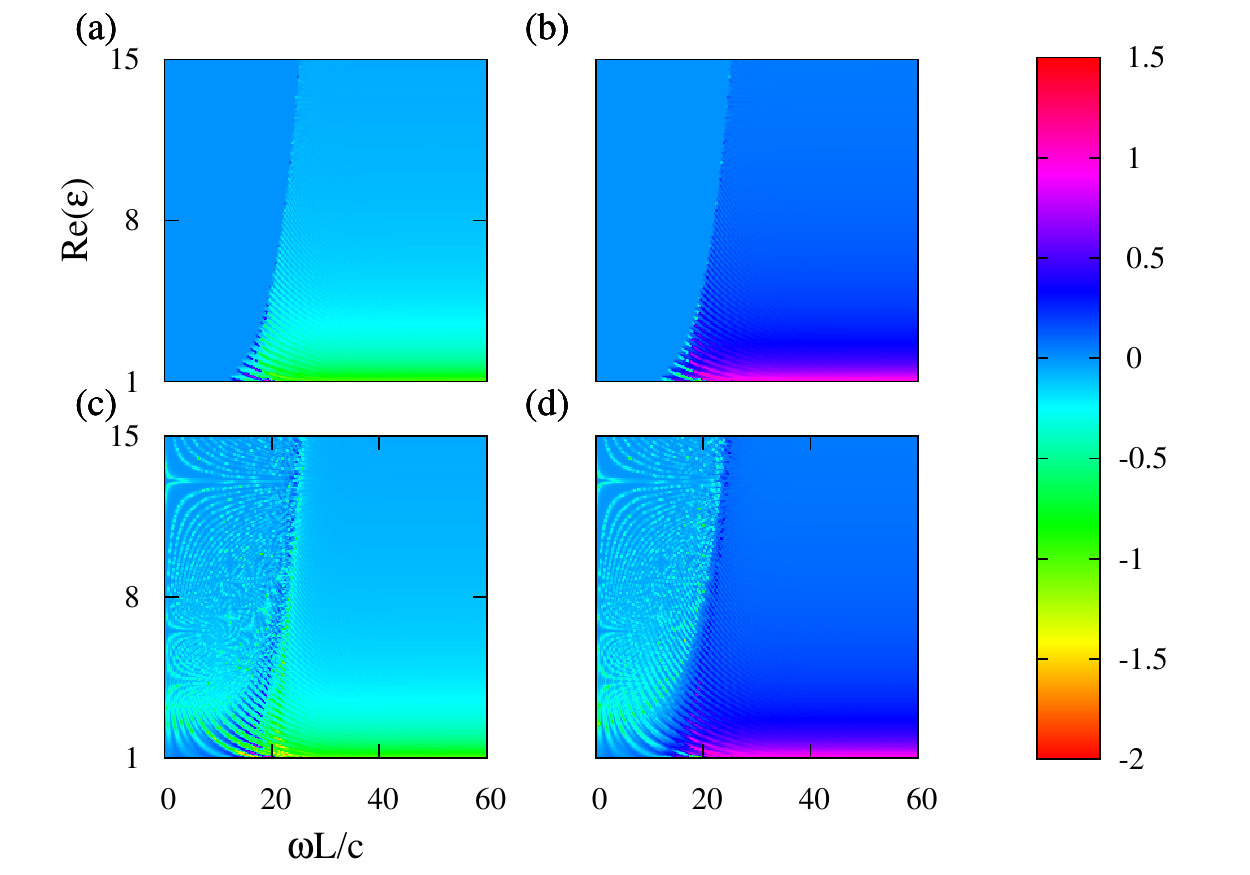}}
\caption{(Color online) Same as Fig. \ref{0-1} but for $Re(\varepsilon)=[1,15]$; $\lambda_1$ (a), $\lambda_2$ (b), 
$\lambda_3$ (c), and $\lambda_4$ (d). As the $Re(\varepsilon)$ increases, the critical frequency $\omega_c$ is 
blueshifted that results in a larger region of simultaneous symmetric (unimodular) and symmetry-broken sets of eigenvalues.}
\label{1-15}
\end{figure} 
Fig. \ref{0-1} and Fig. \ref{1-15} show the maps of the magnitudes of eigenvalues $\lambda_m$, $m=1,2,3,4$, in log10 scale, which are obtained for Case 2 
from Eqs. (\ref{eq:Eq8a}) and (\ref{eq:Eq8b}), in wide ranges of variation in 
$Re(\varepsilon)$ and $\omega$. It is clearly seen that the symmetry of eigenvalues $\lambda_{3,4}$ is broken even 
at small frequencies. Thus, there is a large region in $(\omega{L}/c, Re(\varepsilon))$-plane, 
where ${\cal PT}$-symmetric and ${\cal PT}$-broken sets of eigenvalues may co-exist. 
Starting from $\omega=\omega_c$, the 
set of eigenvalues $\lambda_{1,2}$ also experiences spontaneous symmetry breaking, so the mixing does not exist anymore. 

As shown in Fig. \ref{0-1}(a,b), 
the value of $\omega=\omega_c$ and the extent to which the symmetry is broken strongly depend 
on $Re(\varepsilon)$. For instance, we obtain $\mbox{min}(\omega_c{L}/c)=5$ in the vicinity of $Re(\varepsilon) = 0.5$. 
For the 
set of eigenvalues $\lambda_{3,4}$,
a strong deviation from the unimodular case is observed in Fig. \ref{0-1}(c,d) even at very small values of $Re(\varepsilon)$. 
In particular, a strong anomaly of $\lambda_{3,4}$ occurs nearly at $0<Re(\varepsilon)<0.5$ and $2<\omega{L}/c<4$.
Hence, ${\cal PT}$-symmetric and ${\cal PT}$-broken eigenvalues may co-exist also in ENZ regime, i.e., in the close vicinity of 
$Re(\varepsilon)=0$. 
It is noteworthy that the behaviours of eigenvalues in Case 1 and Case 2 at $0<Re(\varepsilon)<1$ are very different.  
In Fig. \ref{1-15}, the boundary between the regions with the mixed phase for eigenvalues (at smaller $\omega{L}/c$) and with 
all symmetry-broken 
eigenvalues (at larger $\omega{L}/c$) is clearly seen. Its location can be controlled by variations in 
$Re(\varepsilon)$, while $Im(\varepsilon)$ is fixed. 
A detailed investigation of these scenarios will be a subject of our future research.

\begin{figure}[t!]
\centerline{\includegraphics[scale=1.2]{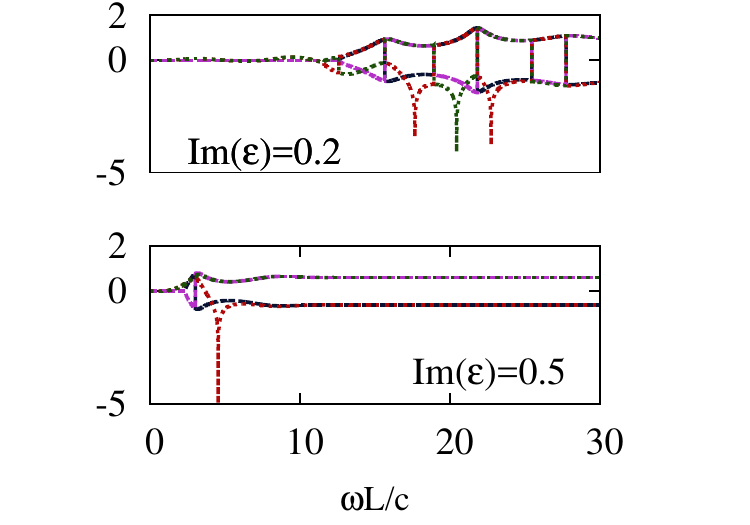}}
\caption{(Color online) The magnitudes of the four eigenvalues (in log10 scale) are plotted against normalized frequency, $\omega{L}/c$, for $Re(\varepsilon)=1$ and $Im(\varepsilon)=0.2$ (top), $Im(\varepsilon)=0.5$ (bottom).
}
\label{3a}
\end{figure}

\begin{figure}[t!]
\centerline{\includegraphics[scale=1.2]{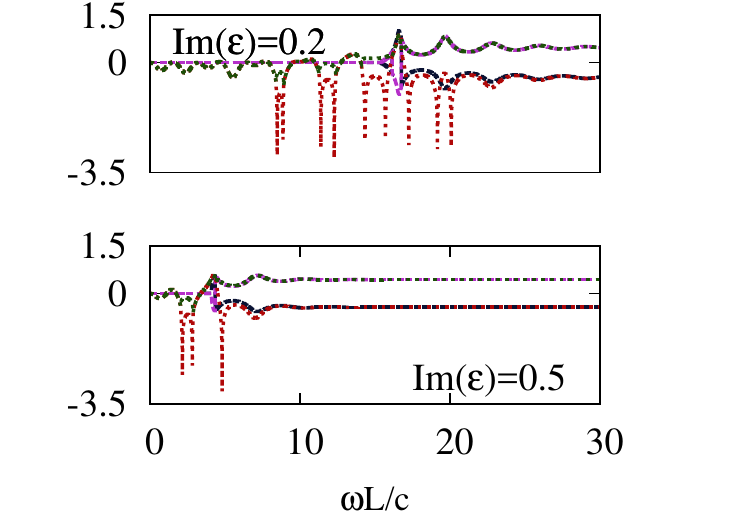}}
\caption{(Color online) The magnitudes of the four eigenvalues (in log10 scale) are plotted against normalized frequency, $\omega{L}/c$, for $Re(\varepsilon)=2$ and $Im(\varepsilon)=0.2$ (top), $Im(\varepsilon)=0.5$ (bottom).
}
\label{3a-2}
\end{figure}

Fig. \ref{3a} and Fig. \ref{3a-2} display the magnitudes for eigenvalues $\lambda_1$, $\lambda_2$, $\lambda_3$ and $\lambda_4$ in log10 scale at varying $\omega{L}/c$ for $Re(\varepsilon)=1$ and $Re(\varepsilon)=2$, respectively, for two different values of $Im(\varepsilon)$. 
The location of the critical frequency for $\lambda_1$ and $\lambda_2$, 
$\omega=\omega_c$, and, hence, width of the $\omega$-range, in which the mixed phase is achieved, are strongly affected by $Im(\varepsilon)$.
As the value of $Im(\varepsilon)$ increases, the critical frequency is 
redshifted, whereas an increase of $Re(\varepsilon)$ results in 
blueshift of $\omega_c$. 
At the same time, $Re(\varepsilon)$ weakly affects the width of mixing range. On the other hand, $Re(\varepsilon)$ can strongly affect the magnitudes of 
$\lambda_{3}$ and $\lambda_{4}$. 
${\cal PT}$-symmetric and ${\cal PT}$-broken sets of eigenvalues can co-exist even at $Re(\varepsilon) = 1$, although the difference between them is rather weak in this case.
Since breaking symmetry at $\omega=\omega_c$ is a general property of $\lambda_{1}$ and $\lambda_{2}$, 
it also occurs in the ENZ regime, e.g. at $Re(\varepsilon) = 0.02$ (not shown). 
However, in this regime, the difference between $\lambda_1$ and $\lambda_2$ at $\omega>\omega_c$ is weak, whereas $\lambda_3$ and $\lambda_4$ may significantly deviate from the unimodular case in a wide range of $Im(\varepsilon)$ variation.

Next, let us add the component $R_0$ to the right end-face of the bilayer, so we now have the configuration $MR_0$. Then, we obtain exactly the same physical scenario for configuration $L_{0}M$, 
while the roles of right and left side illuminations are interchanged. The $\mathbb{S}$-matrix is now given as:  
\begin{ceqn}
\begin{align}
\mathbb{S}(\omega) = \begin{pmatrix} 0 & 0 & r_R & t \cr 0 & r_L & t & 0 \cr r_R & t & 0 & 0 \cr t & 0 & 0 & r_L \end{pmatrix}.
\end{align}
\end{ceqn}
It has four eigenvalues, which may yield the mixed phase: 
\begin{subequations}
\begin{ceqn}
\begin{align}
\lambda_{1,2} = \frac{1}{2} \Big \{  (r_R+r_L) \pm \sqrt{(r_R-r_L)^2 +4t^2} \Big \}, 
\label{eq:Eq10a}  \\
\lambda_{3,4} = \frac{1}{2} \Big \{  (r_L-r_R) \pm \sqrt{(r_R+r_L)^2 +4t^2} \Big \}.
\label{eq:Eq10b}
\end{align}
\end{ceqn}
\end{subequations} 

\begin{table}[h!]
\caption{Number of eigenvalues and ability of mixing eigenvalue phases for different configurations with a different number and properties of the individual components.} 
\begin{center}
\begin{tabular}{ l | c | c | c } 
 \hline \hline
 & eigenvalues &  components & mixed phase \\ \hline
 ${\bf M}$ & 2 & 1 & \xmark \\ \hline 
 {$\bf L_0$}{$\bf M$}{$\bf R_0$} & 4 & 3 & \xmark \\ \hline 
 {$\bf L_0$}{$\bf M$} or {$\bf M$}{$\bf R_0$} & 4 & 2 & \cmark \\ \hline
 {$\bf L_1$}{$\bf M$}{$\bf R_1$} & 4 & 3 & \xmark \\ \hline
 {$\bf L_2$}{$\bf M$}{$\bf R_2$} & 4 & 3 & \xmark \\ \hline
 {$\bf L_1$}{$\bf M$} or {$\bf M$}{$\bf R_1$} & 4 & 2 & \xmark \\ \hline
 {$\bf L_2$}{$\bf M$} or {$\bf M$}{$\bf R_2$} & 4 & 2 & \cmark \\ \hline
 {$\bf L_0$}{$\bf M$}{$\bf R_1$} or {$\bf L_1$}{$\bf M$}{$\bf R_0$} & 4 & 3 & \cmark \\ \hline
 {$\bf L_0$}{$\bf M$}{$\bf R_2$} or {$\bf L_2$}{$\bf M$}{$\bf R_0$} & 4 & 3 & \xmark \\ \hline  
 {$\bf L_1$}{$\bf M$}{$\bf R_2$} or {$\bf L_2$}{$\bf M$}{$\bf R_1$} & 4 & 3 & \cmark \\ \hline \hline  
\end{tabular}
\end{center}
\label{prop}
\end{table}

Similarly to Case 1, we can obtain the sets of eigenvalues given by Eqs. (\ref{eq:Eq8a}), (\ref{eq:Eq8b}) and by Eqs. (\ref{eq:Eq10a}), (\ref{eq:Eq10b}), 
in different configurations. The $\mathbb{S}$-matrices of these configurations should be equivalent to the $\mathbb{S}$-matrices 
of the $L_0M$ and $MR_0$ cases, respectively, from which they must be obtainable via unitary transformations to 
yield the same eigenvalues. From the obtained results, 
the first set of eigenvalues [Eqs. (\ref{eq:Eq8a}), (\ref{eq:Eq8b})], 
can be accessed in the configurations $L_2M$, $L_0MR_1$, and $L_1MR_2$, whereas the second set of eigenvalues [Eqs. (\ref{eq:Eq10a}), \ref{eq:Eq10b})] does in 
$MR_2$, $L_1MR_0$, and $L_2MR_1$.
The simplest configurations for Case 2 are $L_0M$, $MR_0$, $L_2M$, and $MR_2$.
They only need two components, i.e., a loss/gain bilayer and one PCC.  

The properties of the studied configurations are summarized in Table \ref{prop}. One can see therein which features are required for obtaining the mixing phase for eigenvalues, and which are for keeping only ${\cal PT}$-symmetric eigenvalues within a certain frequency range. 
Structures with two components - one loss/gain bilayer and one PCC with peculiar properties - can be sufficient for obtaining the mixing phase, whereas those with three components do not always lead to it. 
Whether the mixing of the phase is achieved or not depends on the polarization conversion scenario at the end-faces, so the role of PCC(s) is evident. If it is achieved, it occurs in a rather wide frequency range.  
It is noticeable that despite having different end-faces $all$ of the studied configurations still enable the sets of symmetric eigenvalues.  
On the other hand, the effect exerted by one PCC can be compensated by that of the other, so phase mixing is not achieved in some three-component 
configurations. This case can be useful for the separation of two processes in one structure and, therefore, is promising for multifunctional operation. 

\section*{Discussion}
The main goal of this study was to show that broadband mixing is possible, and, moreover, it may be achieved in photonic heterostructues with a one-dimensional gain/loss bilayer.  
We demonstrated a way to broadband mixing of ${\cal PT}$-symmetric and ${\cal PT}$-broken phases for eigenvalues in the structures with four channels and with a one-dimensional loss/gain component. 
As far as symmetric and broken phases show different properties related to flux conservation, amplification, and attenuation, it is expected that the broadband mixing of these phases may open new routes to efficient selective manipulation by electromagnetic radiation, including advanced regimes of directional selectivity, enhancement, and absorption.  
Earlier, such a mixing has been expected to occur for physical dimensions higher than one. The obtained results show that a 
two-dimensional loss/gain medium and, moreover, anisotropy are not required for the mixing.  
We have analytically derived the eigenvalues for different configurations of photonic heterostructures consisting of a bilayer of one-dimensional loss/gain 
medium and PCC(s), and proved the possibility of obtaining 
the mixed phase for eigenvalues in a wide but limited frequency range, while its width depends on both the real and the imaginary part 
of the permittivity of the loss/gain component. Therefore, the wideband phase mixing is a very general effect whose existence does not need 
any special adjustment of the parameters, although its appearance can be strongly sensitive to the parameter choice. 
While the principal possibility of obtaining four channels by replacing a two-dimensional loss/gain medium with a one-dimensional medium combined with a PCC could be expected, the obtained results indicate that this case can be achieved in different permittivity ranges for multiple configurations, which are distinguished due to the presence and properties of the PCC(s).  
The utilized model based on $\mathbb{S}$-matrix formalism  
properly describes a wide variety of photonic heterostructures with a 1D loss/gain component and their ${\cal PT}$-related properties.    
It is noteworthy that the ${\cal PT}$-symmetric eigenvalues set may exist for all of the considered configurations, regardless of whether the loss/gain component 
is end-faced with one or two PCCs, or not end-faced at all. 
According to the goals of this study, we clarified, based on the obtained results, which properties of PCCs are necessary and which are sufficient in different transmission/reflection scenarios. We showed that a two-component heterostructure can be sufficient to obtain wideband phase mixing, provided that the PCC(s) show the suitable properties. On the other hand, we detected such three-component heterostructures that keep immunity against phase mixing, in spite of containing a PCC that may lead to the mixing in other, even simpler configurations.  
Design of PCCs enabling the desired polarization manipulation will be one of the next steps.  
A further study of the peculiarities of different ranges of material parameters is planned. 
The difference between the cases with and without phase mixing for eigenvalues is especially intriguing for ultralow-permittivity regime, in which the mixing is possible even in close vicinity of $Re(\varepsilon)=0$. Thus, the consideration of the studied physical features in connection with recent advances in theory and applications of ENZ materials and specific coupling and transmission regimes realized with their use is a promising topic for future research. 

\section*{Methods}
The standard $\mathbb{S}$-matrix formalism, which is commonly used in quantum mechanics, has been used to mathematically express transmission and reflection properties of different configurations. The eigenvalues of the $\mathbb{S}$-matrix were analyzed analytically. 
The values of $r_L$, $r_R$ and $t$ were calculated by applying the conditions of continuity for the tangential components of electromagnetic field at the boundaries of the structural components, so the unknown coefficients in the general formulas for the field components are introduced unambiguously. 
The eigenvalues were calculated for various configurations via a simple custom-made Fortran code and the results are plotted via GNUPLOT. Both of these procedures were performed on a standard laptop that works on an Ubuntu operating system.

\section*{Acknowledgements}

This work is supported by the projects DPT-HAMIT and NATO-SET-193. One of the
authors (E.O.) also acknowledges partial support from the Turkish Academy of
Sciences. A.E.S. thanks the National Science Centre of Poland for support under the
project MetaSel DEC-2015/17/B/ST3/00118. Work at Ames Laboratory was partially supported by the U.S. Department of Energy, Office of Basic Energy Science, 
Division of Materials Sciences and Engineering (Ames Laboratory is operated for the U.S. Department of Energy by Iowa State University under 
Contract No. DE-AC02-07CH11358). The European Research Council under ERC Advanced Grant No. 320081 (PHOTOMETA) supported work at FORTH. This work is origally published in Scientific Reports {\bf [Sci Rep. 2017; 7: 15504.]}.

\end{document}